\documentclass[aps,twocolumn,prd,showpacs,showkeys,preprintnumbers,superscriptaddress,bibnotes,floatfix,longbibliography,notitlepage,nofootinbib,groupedaddress,reprint]{revtex4-2}

\usepackage[english]{babel}
\usepackage{amsfonts}

\DeclareSymbolFontAlphabet{\mathbb}{AMSb}

\usepackage{enumitem}
\usepackage{braket}
\usepackage{amsmath,amssymb}
\usepackage{txfonts}
\usepackage{comment}
\usepackage{bm}
\usepackage{mathtools}
\usepackage{makecell}

\usepackage[usenames, dvipsnames, svgnames, table]{xcolor}
\usepackage[colorlinks=true, citecolor=Blue,
            linkcolor=BrickRed, urlcolor=ForestGreen]{hyperref}
\usepackage[capitalise]{cleveref}

\usepackage{graphicx}
\usepackage{xfrac}
\usepackage{ulem}
\usepackage{cancel}

\usepackage[
raggedright
]{sidecap}   
\sidecaptionvpos{figure}{c} 

\begin{document}

\title{Misalignment and mode mismatch error signals for higher-order Hermite-Gauss modes from two sensing schemes}

\overfullrule 0pt 
\parskip 0pt
\hyphenpenalty 9999

\author{Liu Tao}
\email{liu.tao@ligo.org} 
\affiliation{University of Florida, 2001 Museum Road, Gainesville, Florida 32611, USA}

\author{Anna C.~Green}
\affiliation{Nikhef, Science Park 105, 1098, XG Amsterdam, The Netherlands}

\author{Paul Fulda}
\affiliation{University of Florida, 2001 Museum Road, Gainesville, Florida 32611, USA}

\begin{abstract}
The locking of lasers to optical cavities is ubiquitously required in the field of precision interferometry such as Advanced LIGO to yield optimal sensitivity. Using higher-order Hermite-Gauss (HG) modes for the main interferometer beam has been a topic of recent study, due to their potential for reducing thermal noise of the test masses. It has been shown however that higher-order HG modes are more susceptible to coupling losses into optical cavities: the misalignment and mode mismatch induced power losses scale as $2n+1$ and $n^{2}+n+1$ respectively with $n$ being the mode index. In this paper we calculate analytically for the first time the alignment and mode mismatch sensing signals for arbitrary higher-order HG modes with both the traditional sensing schemes (using Gouy phase telescopes and quadrant photodetectors) and the more recently proposed radio-frequency jitter-based sensing schemes (using only single element photodiodes). We show that the sensing signals and also the signal-to-shot noise ratios for higher-order HG modes are larger than for the fundamental mode. In particular, the alignment and mode mismatch sensing signals in the traditional sensing schemes scale approximately as $\sqrt{n}$ and $n$ respectively, whereas in the jitter-based sensing schemes they scale exactly as $2n+1$ and $n^{2}+n+1$, respectively, which exactly matches the decrease in their respective tolerances. This potentially mitigates the downside of higher-order HG modes for their suffering from excessive misalignment and mode-mismatch induced power losses. 
\end{abstract}

\maketitle

\section{Introduction}
The sensitivity of all leading gravitational wave detectors such as Advanced LIGO and Advanced Virgo~\cite{aLIGO,PhysRevD.102.062003,AdVirgo} at signal frequencies around 100 Hz is limited by the thermal noise of the test-mass optics. This is also likely to be the case for next generation detectors such as Cosmic Explorer and Einstein Telescope~\cite{CE,ET}. To reduce this noise and thus obtain better detector sensitivity, the idea of replacing the currently used fundamental Gaussian laser beam with higher-order optical modes such as Laguerre-Gauss (LG) modes or Hermite-Gauss (HG) modes has been proposed. Higher-order modes with more uniform intensity distributions than the fundamental Gaussian beam can better ``average out'' the effects of this thermal noise~\cite{Mours_2006, Vinet_2007}. Research into the potential use of higher-order $\mathrm{LG}_{3,3}$ mode carried out with numerical simulations and tabletop experiments~\cite{PhysRevD.79.122002, PhysRevD.82.012002} has shown that the surface figure imperfections present even in state-of-the-art mirrors will cause significant impurity and losses for the $\mathrm{LG}_{3,3}$ mode in realistic, high finesse cavities~\cite{PhysRevD.84.102002, PhysRevD.84.102001, Sorazu}. However, it has also been shown~\cite{tao2020higherorder, PhysRevD.103.042008} that higher-order HG modes such as $\mathrm{HG}_{3,3}$ mode can be made almost as robust as the currently used $\mathrm{HG}_{0,0}$ against mirror surface deformations with the deliberate addition of astigmatism to the test masses.

A. Jones et al.~\cite{Jones_2020} and the authors of this paper~\cite{Tao:21} have shown however the misalignment and mode mismatch induced power coupling losses increase with the mode order. As a result, higher-order $\text{HG}$ modes are more sensitive to beam distortions such as misalignment and mode mismatch than the fundamental mode. However, in this paper we show that higher-order $\text{HG}$ modes also have stronger alignment and mode mismatch sensing signals and thus better sensing and control capabilities. This is essential for the implementation of higher order HG modes in future gravitational wave detectors for their thermal noise benefits, as well as in many other areas that utilize the beneficial higher order spatial laser modes, such as in multimode optical
quantum information systems~\cite{PhysRevLett.98.083602} and a variety of microscopy-related systems for high resolution imaging~\cite{PhysRevLett.110.043601}. We demonstrate this result by calculating the so-called relative sensing gain, which is defined as the error signal ratio of higher-order mode compared against the fundamental mode. In this paper we will investigate both analytically and through \textsc{Finesse}~\cite{Freise_2004, finesse, pykat,brown2020pykat} simulations the alignment and mode mismatch sensing gains as functions of $\text{HG}$ mode index for a single symmetric Fabry–Perot cavity in different detection schemes. 

\begin{figure*}[!htb]
    \centering
        \includegraphics[width=\textwidth]  {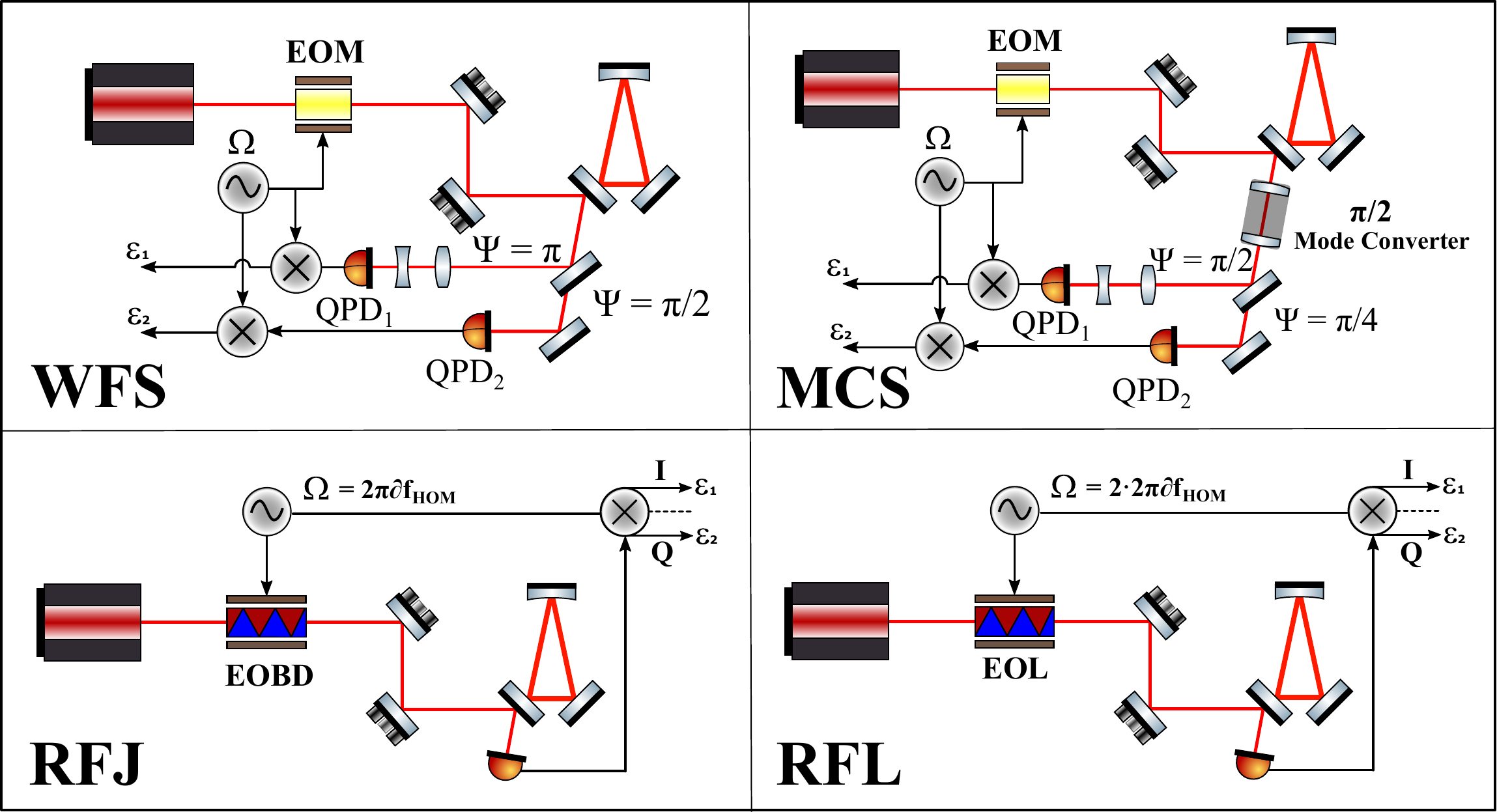}
        \caption{\normalsize Illustration of various alignment and mode mismatch sensing schemes considered in this paper. The traditional sensing schemes are on the top row, and the more recently proposed jitter-based sensing schemes are shown on the bottom.}
        \label{fig:schemes}
\end{figure*}

We consider the traditional heterodyne wavefront sensing (WFS) scheme~\cite{Morrison1:94, Morrison2:94} and the more recently proposed radio-frequency jitter modulation sensing (RFJ) scheme~\cite{Fulda:17} for the alignment sensing, and the mode-converter sensing (MCS) scheme~\cite{Maga_a_Sandoval_2019} and radio-frequency lens modulation sensing (RFL) scheme~\cite{Fulda:17, Ciobanu:20, dcc-LIGO-G2001619, dcc-LIGO-G2001575} for mode mismatch sensing, as shown in Fig.~\ref{fig:schemes}.
Specifically, WFS uses two split photodetectors separated by $\frac{\pi}{2}$ radians Gouy phase for the two orthogonal misalignment degrees of freedom per axis, namely tilt and lateral displacement. RFJ for alignment sensing on the other hand makes use of an electro-optic beam deflector (EOBD) to impose RF jitter sidebands separated from the carrier frequency by the higher-order mode difference frequency of the optical cavity. Demodulating the beat signal between the RF jitter modulation-induced offset mode sidebands and the misalignment-induced carrier frequency offset modes on a single element photodiode in orthogonal demodulation phases (i.e. separated by $\frac{\pi}{2}$ radians) results in linear error signals for both tilt and translation of the input beam with respect to the cavity axis. 
For MCS we use an astigmatic mode-converter and two $45^\circ$-rotated quadrant photodetectors separated by $\frac{\pi}{4}$ radians Gouy phase to sense the two orthogonal mode mismatch degrees of freedom, namely waist size and waist position mismatch. 
RFL for the mode mismatch sensing on the other hand uses an electro-optic lens device (EOL) to produce RF defocus sidebands separated from the carrier frequency by \textit{twice} the higher-order mode difference frequency of the optical cavity to simultaneously extract the full mode mismatch error signals in the orthogonal demodulation phases from a single photodiode. 
We show that in all the aforementioned sensing schemes, higher-order HG modes always have stronger sensing signals compared against the fundamental mode. In particular, RFJ/L schemes show an increase in the sensing gains and the signal to shot noise ratio that exactly matches the decrease in the corresponding tolerance~\cite{Jones_2020, Tao:21}, which potentially mitigates the downside of higher-order HG modes for their suffering from excessive misalignment and mode-mismatch induced power losses in sensing-noise-limited interferometers. We also conduct the corresponding \textsc{Finesse} simulations for comparison against the analytical results for the sensing gains. The \textsc{Finesse} results agree extremely well with the analytical models.

This paper is structured as follows: In Section~\ref{sec:alignment} and~\ref{sec:modemismatch} we present a step by step theoretical derivation of the alignment and mode mismatch sensing signals in symmetric cavities for arbitrary $\text{HG}$ modes, respectively. We also include a comparison with the corresponding \textsc{Finesse} simulation results. We report our conclusions and discussions in Section~\ref{sec:conclusion}. 

\begin{figure}[htbp]
    \centering
    \includegraphics[width=\linewidth]{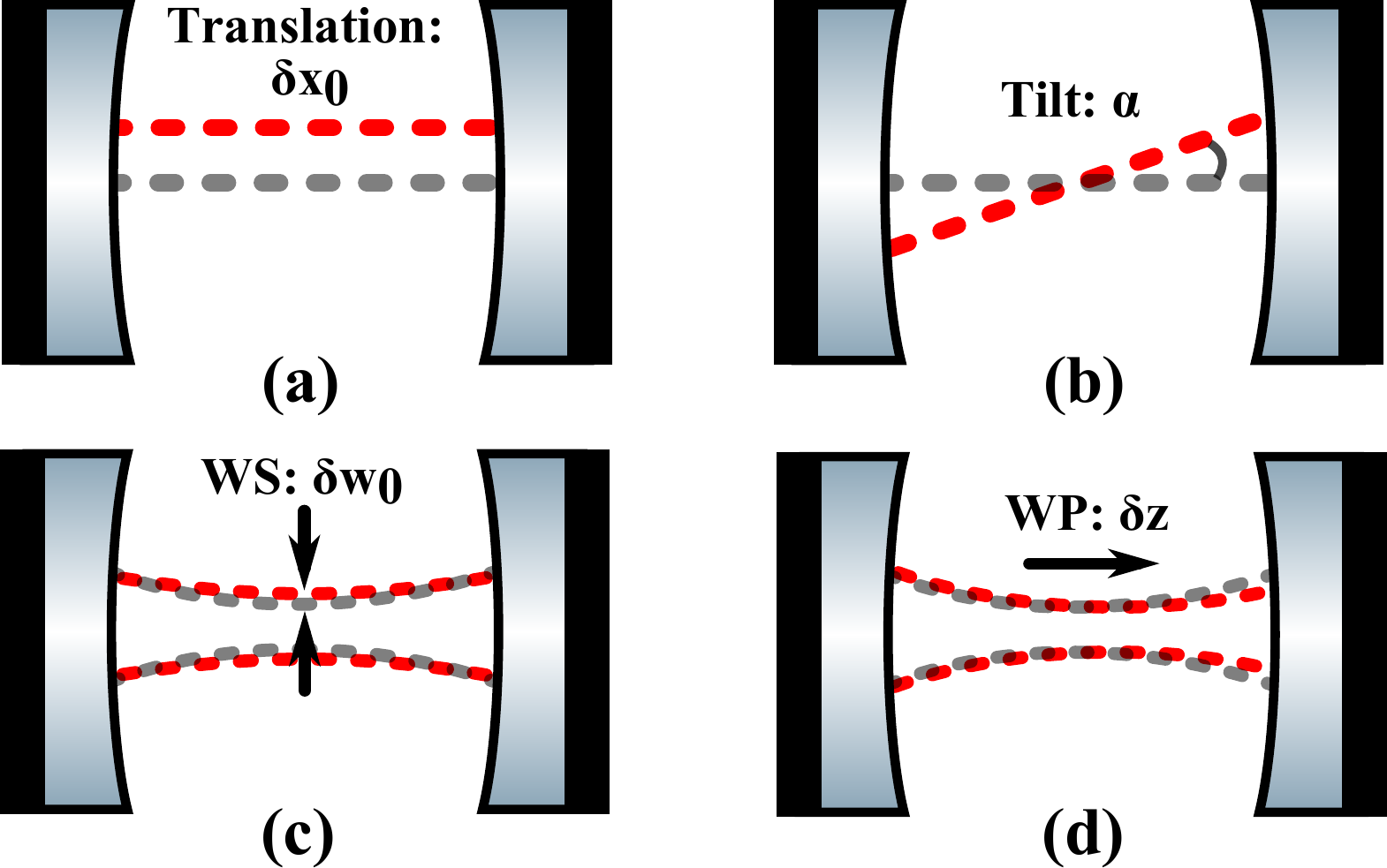}
    \caption{Illustration of misalignment (top) and mode mismatch (bottom) between the input laser mode (red) and the cavity eigenmode (grey). Top, the left panel shows lateral translation of $\delta x_{0}$ and the right panel shows tilt of $\alpha$. Bottom, the left panel shows the waist size mismatch of amount $\delta w_{0}$ and the right shows the waist position mismatch of amount $\delta z$.}
    \label{fig:alignmentModeMismatch}
\end{figure}

\section{Alignment Sensing}
\label{sec:alignment}
In the following two subsections, we will calculate the alignment sensing error signals in the traditional wavefront sensing scheme (WFS)~\cite{Morrison1:94, Morrison2:94} and the more recently proposed radio-frequency jitter modulation sensing scheme (RFJ)~\cite{Fulda:17} with an arbitrary higher-order HG mode as the carrier.

The general expression for a HG mode propagating along the $z$ axis is~\cite{Bond2017}
\begin{equation}
\mathcal{U}_{n, m}(x, y, z)=\mathcal{U}_{n}(x, z) \mathcal{U}_{m}(y, z)
\label{equ:HGnm}
\end{equation}
with 
\begin{equation}
\begin{aligned}
\mathcal{U}_{n}(x, z)&=\left(\frac{2}{\pi}\right)^{1 / 4}\left(\frac{\exp (\mathrm{i}(2 n+1) \Psi(z))}{2^{n} n ! w(z)}\right)^{1 / 2} \\
& \times H_{n}\left(\frac{\sqrt{2} x}{w(z)}\right) \exp \left(-\mathrm{i} \frac{k x^{2}}{2 R_{c}(z)}-\frac{x^{2}}{w^{2}(z)}\right) \,,
\end{aligned}
\label{equ:HGn0}
\end{equation}
where $\Psi(z) = \arctan \left(\frac{z-z_{0}}{z_{R}}\right)$ is the Gouy phase with $z_{R} = \frac{\pi w_{0}^2}{\lambda}$ being the Rayleigh range. $k$ is the wavenumber, $\lambda$ is the wavelength, $w(z)$ is the beam radius and $R_{c}(z)$ is the wavefront radius of curvature. They are related to the beam waist size $w_{0}$ and beam waist position $z_{0}$ via
\begin{equation}
\begin{aligned}
w(z)&=w_{0}+\sqrt{1+\left(\frac{z-z_{0}}{z_{R}}\right)^{2}} \\
R_{c}(z)&=z-z_{0}+\frac{z_{R}^{2}}{z-z_{0}}
\end{aligned}
\end{equation}
The single mode electric field before entering the cavity can thus be written as 
\begin{equation}
E(x, y, z)=E_{0} \mathcal{U}_{n,m}(x, y, z) e^{\mathrm{i}\left(\omega t-k z\right)}
\label{equ:initialbeam}
\end{equation}
where $E_{0}$ is the initial field amplitude and $\omega$ is the field angular frequency. 

For alignment sensing, as illustrated in the top panels of Fig.~\ref{fig:alignmentModeMismatch}, since HG modes are separable in $x$ and $y$, and therefore the sensing gains for misalignment in one axis are independent of the mode index in the orthogonal axis, one can always consider the single-axis behaviour, such as a misalignment in the $xz$ plane for the $\mathrm{HG}_{n,0}$ mode, without loss of generality~\cite{Tao:21}. By symmetry the same arguments can be applied to the misalignment in the $yz$ plane for the $\mathrm{HG}_{0,m}$ mode and $\mathrm{HG}_{n,m}$ modes in general.

In this Section we consider a generic $\mathrm{HG}_{n,0}$ mode with the misalignment of the beam occurring in the $xz$ plane at the center of a symmetric cavity, which we make coincident with the origin of our coordinate system, i.e. $z_{0} = 0$. We can thus simplify the initial beam as
\begin{equation}
E(x, z)=E_{0} \mathcal{U}_{n}(x, z) e^{\mathrm{i}\omega t}
\label{equ:alignmentbeam}
\end{equation}

\subsection{Alignment Sensing: Wavefront Sensing (WFS)}
In WFS scheme, we make use of two quadrant photodetectors in reflection of the cavity away from the cavity waist by $\pi$ and $\pi/2$ Gouy phases for the tilt and lateral offset degrees of freedom, respectively. 

\subsubsection{WFS: tilt}
Let us look at the tilt degree of freedom first. We apply a phase modulation with modulation index $m$ at a frequency $\Omega$ to the carrier field in the $\text{HG}_{n,0}$ mode. Keeping the first order sidebands only, the beam~\ref{equ:alignmentbeam} becomes
\begin{equation}
E=E_{0} \mathcal{U}_{n} e^{\mathrm{i} \omega t}\left(1+\mathrm{i} \frac{m}{2}\left(e^{-\mathrm{i} \Omega t}+e^{\mathrm{i} \Omega t}\right)\right)
\label{equ:wfs_tilt_modulation}
\end{equation}
According to the results in paper~\cite{Tao:21}, upon application of a misalignment angle $\alpha$ between the input beam axis and the cavity optical axis about the cavity waist, the field at the cavity input mirror, up to the first order, becomes
\begin{equation}
\begin{aligned}
\mathcal{U}^{\alpha}_n&\approx \mathcal{U}_{n}
+ \mathrm{i} \frac{\alpha}{\Theta} \bigg(\sqrt{n+1} \mathcal{U}_{n+1}e^{-\mathrm{i}\Psi} + \sqrt{n} \mathcal{U}_{n-1} e^{\mathrm{i}\Psi} \bigg)
\end{aligned}
\label{equ:tilted}
\end{equation}
where $\Theta = \frac{\lambda}{\pi w_{0}}$ is the far-field divergence angle and $\Psi$ is the accumulated Gouy phase from the input mirror to the cavity waist. The input modulated beam from Eq.~\ref{equ:wfs_tilt_modulation}, now reads
\begin{equation}
\begin{aligned}
E=E_{0} &\left(\mathcal{U}_{n} + \frac{\mathrm{i} \alpha}{\Theta} \left(\sqrt{n+1} \mathcal{U}_{n+1}e^{-\mathrm{i}\Psi} + \sqrt{n} \mathcal{U}_{n-1} e^{\mathrm{i}\Psi} \right) \right) e^{\mathrm{i}\omega t} \\
&\times \left(1+\mathrm{i} \frac{m}{2}\left(e^{-\mathrm{i} \Omega t}+e^{\mathrm{i} \Omega t}\right)\right)
\end{aligned}
\label{equ:input}
\end{equation}
In order to detect an alignment error signal in reflection of the cavity, we must consider the reflected field from the cavity $\mathrm{E}_\mathrm{refl}$. In reflection, each term in Eq.~\ref{equ:input} is multiplied by the cavity complex reflectance function $F(\omega,n)$, where n is the mode index. The complex reflectance in general is given by
\begin{equation}
F(\omega, n)=r_{1}-\frac{r_{2} t_{1}^{2} \exp \left(-\mathrm{i}\left(D \frac{\omega}{c}+(n+1)\Psi_{r t}\right)\right)}{1-r_{1} r_{2} \exp \left(-\mathrm{i}\left(D \frac{\omega}{c}+(n+1)\Psi_{r t}\right)\right)}
\label{equ:fwn}
\end{equation}
where $r_{1}$, $t_{1}$, $r_{2}$, and $t_{2}$ are the amplitude reflectivities and transmissivities of the ITM and ETM, respectively; $D$ is the exact round-trip length of the cavity; and $\Psi_{r t}$ the Gouy phase accumulated on one round-trip path inside the cavity. In general the cavity complex reflectance function $F(\omega,n)$ is complicated. However, for a high-finesse, completely over-coupled cavity that is geometrically stable, all reflectivity coefficients are real, and either equal to 1 for non-resonant field components, or -1 for resonant field components. For example, the cavities in Advanced LIGO have similar properties and so we can make the assumptions throughout to make the analytical results more intuitively understandable. As a result, we have $F(\omega, n) = -1$ while the reflectance functions for the non-resonant field components, such as $F(\omega, n\pm1)$, $F(\omega\pm\Omega, n)$, and $F(\omega\pm\Omega, n\pm1)$, are all 1.

The reflected field $\mathrm{E}_{\text{refl}}$ with the above assumption is
\begin{equation}
\begin{aligned}
\text E_\mathrm{refl} &= \Bigg( -\mathcal{U}_{n} + \frac{\mathrm{i} m}{2}\bigg(\mathcal{U}_{n} e^{\mathrm{i} \Omega t} +  \mathcal{U}_{n} e^{-\mathrm{i} \Omega t}\bigg) - \frac{\mathrm{i} \alpha}{\Theta}\bigg( \mathcal{U}_{n+1} \sqrt{n+1} + \\
&\mathcal{U}_{n-1} \sqrt{n} \bigg) +\frac{\alpha}{\Theta} \frac{m}{2} \bigg( \mathcal{U}_{n+1} \sqrt{n+1}  e^{\mathrm{i}\Omega t} +  \mathcal{U}_{n+1} \sqrt{n+1}  e^{-\mathrm{i}\Omega t} \\
&+  \mathcal{U}_{n-1} \sqrt{n}  e^{\mathrm{i}\Omega t} +  \mathcal{U}_{n-1} \sqrt{n} e^{-\mathrm{i}\Omega t} \bigg) \Bigg)\cdot E_{0} e^{\mathrm{i} \omega t} 
\end{aligned}
\end{equation}
where we have set the accumulated Gouy phase from the cavity waist to the QPD to be $\pi$. This produces an extra factor of $e^{i \pi \cdot (\pm 1)}=-1$ for the adjacent upper and lower modes scattered from the original mode, as the result of tilt. 

The photocurrent produced by a split photodetector in reflection from the cavity is given by
\begin{equation}
\mathrm{I}_{PD}=\left(\int_{0}^{\infty}d x - \int_{-\infty}^{0} d x \right) E_{\mathrm{refl}} \cdot E_{\mathrm{refl}}^{*}
\end{equation}
for an ideal split photodetector, assuming an appropriate responsivity of 1 $A/W$. 
$E_{\mathrm{refl}}$ and $E_{\mathrm{refl}}^{*}$ have 9 terms each, so there are 81 terms to evaluate in the product $E_{\mathrm{refl}}E_{\mathrm{refl}}^{*}$. However, most terms will have no contribution to the signal. In particular, only terms that are odd in $x$ have contributions to the split photodetector signal. We also know only terms that oscillate with the modulation frequency $\Omega$ contribute to the photodetector signal after demodulation at $\Omega$ and low-pass filtering. The non-vanishing photocurrent is then
\begin{equation}
\begin{aligned}
\mathrm{I}_{PD}&=-2\mathrm{E}_{0}^{2}\frac{\alpha}{\Theta} m \left( e^{\mathrm{i} \Omega t} + e^{-\mathrm{i} \Omega t}\right)\left(\sqrt{n+1}\beta_{n,n+1} + \sqrt{n}\beta_{n,n-1}\right) 
\label{equ:photocurrent}
\end{aligned}
\end{equation}
where $\beta_{n,n+1}$ are the beat coefficients
\begin{equation}
\beta_{n,n+1}=\left(\int_{0}^{\infty}d x - \int_{-\infty}^{0} d x \right) \mathcal{U}_{n}\cdot \mathcal{U}_{n+1}
\label{equ:beatsplit}
\end{equation}
In general evaluating the beat coefficients is complicated, but as n goes large $\beta_{n,n+1}$ approaches a constant value around 0.64. For further details see the appendix~\ref{app:beatcoeffs}.

We may assume the photocurrent signal is converted to a voltage signal by an appropriate trans-impedance stage with a gain of 1 $V/A$. Demodulating the beat signal with a phase $\phi=0$ (a condition often known as in-phase or I-phase demodulation) and removing signal components at $2\Omega$ with a low-pass filter, we obtain the following expression for the tilt error signal measured at $\mathrm{QPD}_{1}$:
\begin{equation}
\mathrm{V}^{\Omega,\, \mathrm{QPD}_{1}}_{\mathrm{WFS}} = -2\mathrm{E}_{0}^{2}\frac{\alpha}{\Theta} m \left(\sqrt{n+1}\beta_{n,n+1} + \sqrt{n}\beta_{n,n-1}\right)
\label{equ:error_tilt_WFS}
\end{equation}
We define the relative sensing gain $\Sigma_{n}$ as the ratio of the error signal for the higher-order mode $\mathrm{HG}_{n,0}$ and the fundamental mode $\mathrm{HG}_{0,0}$
\begin{equation}
\Sigma_{n} \equiv \left(\sqrt{n+1} \beta_{n,n+1} + \sqrt{n} \beta_{n,n-1}\right)/\beta_{0,1}
\label{equ:gain1}
\end{equation}
The sensing gain increases with the mode index.

\subsubsection{WFS: translation.  }
Now let us look at the lateral offset degree of freedom. According to the results in paper~\cite{Tao:21}, the beam after the lateral displacement with respect to the cavity optical axis, at the cavity input mirror becomes
\begin{equation}
\begin{aligned}
\mathcal{U}^{\delta x_{0}}(x, z) \approx \mathcal{U}_{n} + \frac{\delta x_{0}}{w_{0}} \left(\sqrt{n+1} \mathcal{U}_{n+1} e^{-\mathrm{i}\Psi} - \sqrt{n} \mathcal{U}_{n-1} e^{\mathrm{i}\Psi}\right)
\end{aligned}
\label{eqn:translated}
\end{equation}
where $\delta x_{0}$ is the lateral displacement along the $x$ direction. The difference for this path compared earlier is that the accumulated Gouy phase from the cavity waist to the QPD is set to $\pi/2$ instead. This produces an extra factor of $e^{i \frac{\pi}{2} \cdot (\pm 1)} = \pm i$ for the upper and lower adjacent modes scattered from the original mode. We thus have
\begin{equation}
\begin{aligned}
\mathcal{U}^{\delta x_{0}}(x, z) \approx \mathcal{U}_{n} + \mathrm{i} \frac{\delta x_{0}}{w_{0}} \left(\sqrt{n+1} \mathcal{U}_{n+1} e^{-\mathrm{i}\Psi} + \sqrt{n} \mathcal{U}_{n-1} e^{\mathrm{i}\Psi}\right)
\end{aligned}
\end{equation}
Comparing against Eq.~\ref{equ:tilted} (with an additional factor of -1 for the adjacent upper and lower scattered modes from the $\pi$ Gouy phase), we see that the alignment signal calculation for tilt and lateral displacement are essentially the same if one changes the expansion parameter $\frac{\alpha}{\Theta}$ to $-\frac{\delta x_{0}}{w_{0}}$. As a result, we can write down directly the lateral translation error signal from Eq.~\ref{equ:error_tilt_WFS}: 
\begin{equation}
\mathrm{V}^{\Omega,\, \mathrm{QPD}_{2}}_{\mathrm{WFS}} = 2\mathrm{E}_{0}^{2}\frac{\delta x_{0}}{ w_{0}} m \left(\sqrt{n+1} \beta_{n,n+1} + \sqrt{n} \beta_{n,n-1}\right)
\label{equ:error_offset1}
\end{equation}
with the demodulation phase set to zero as well. We also obtain the same sensing gain in this translation path as the tilt case in Eq.~\ref{equ:gain1}.

\begin{figure}[htbp]
    \centering
    \includegraphics[width=\linewidth]{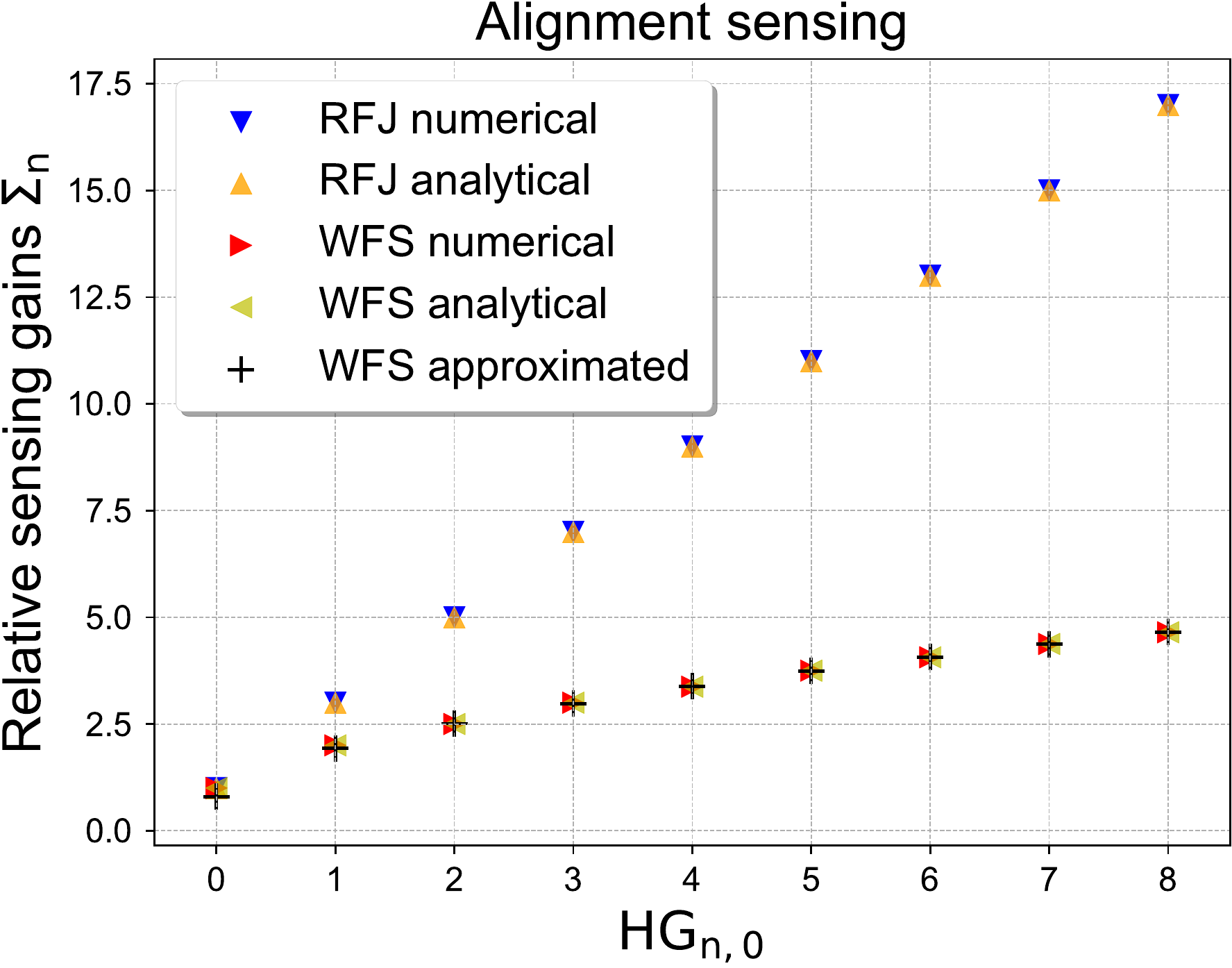}
    \caption{Relative alignment sensing gain $\Sigma_{n}$ in WFS and RFJ schemes for $\mathrm{HG}_{n, 0}$ modes. The analytical sensing gains in Eqs.~\ref{equ:gain1} and~\ref{equ:rfjsensinggain} and the approximation in Eq.~\ref{equ:gain} are included. The corresponding \textsc{Finesse} numerical results are also included for comparison.}
    \label{fig:approx1}
\end{figure}

As shown in Fig.~\ref{fig:beatcoeff1} in the appendix~\ref{app:beatcoeffs}, the beat coefficients converge to around 0.64 as the mode order gets large. We thus can approximate our exact results in Eqs.~\ref{equ:error_tilt_WFS} and~\ref{equ:error_offset1} by replacing the complicated higher-order beat coefficients (i.e. excluding $\beta_{0,1}$) with a constant value 0.64. 
After the  approximation explained above we obtain
\begin{equation}
\Sigma_{n} \approx 0.64\cdot\left(\sqrt{n+1} + \sqrt{n}\right)/\beta_{0,1}
\label{equ:gain}
\end{equation}
As shown in Fig.~\ref{fig:approx1}, the approximation is quite good except for lower mode $\mathrm{HG}_{0, 0}$, which is due to the fact that for lower mode index the beat coefficients, $\beta_{0,1}$ for example, are quite different from our approximation 0.64. We also conduct the corresponding \textsc{Finesse} simulation in black in Fig.~\ref{fig:approx1}. They agree exceptionally well with the exact analytical results in yellow.

We thus conclude that the alignment sensing gain for higher-order mode $\mathrm{HG}_{n,0}$ in this traditional WFS scheme scales approximately as $\sqrt{n} + \sqrt{n+1}$. The calculation is done assuming misalignment in the $xz$ plane for a generic $\text{HG}_{n,0}$ (i.e. 2D), but one can easily generalize it to an arbitrary 3D case since HG modes are separable in $x$ and $y$ axes and any off-axis misalignment can be resolved into a tilt in each axis.

\subsection{Alignment Sensing: Radio-Frequency Jitter sensing (RFJ)}
For the alignment sensing with the RF jitter modulation scheme, we use an electro-optic beam deflector (EOBD) to impose RF jitter modulation sidebands separated from the carrier frequency by the higher-order mode difference frequency of the optical cavity. The scheme involves demodulating the beat signal between the RF jitter modulation-induced offset mode sidebands and the misalignment-induced carrier frequency offset modes on a single-element photodetector in orthogonal demodulation phases. This produces simultaneously linear error signals for the two orthogonal misalignment degrees of freedom. With a single-element photodetector, the beat coefficients between say $\mathrm{HG}_{n,m}$ and $\mathrm{HG}_{n^{\prime},m^{\prime}}$ modes will simply be the Kronecker delta functions
\begin{equation}
\beta_{(n,m),(n^{\prime},m^{\prime})} \equiv \int_{-\infty}^{\infty} \int_{-\infty}^{\infty}\mathrm{d} x \mathrm{d} y \, \mathcal{U}_{n m} \cdot \mathcal{U}_{n^{\prime} m^{\prime}}  = \delta_{n,n^{\prime}} \cdot \delta_{m,m^{\prime}}
\label{equ:beatcoe}
\end{equation}
We will show explicitly with orthogonal demodulation phases this beat signal provides simultaneously linear error signals for both tilt and translation of the two axes.

Similar to the WFS scheme, we consider misalignment in the $xz$ plane for a generic $\text{HG}_{n,0}$ beam propagating along the z axis. The initial field in Eq.~\ref{equ:alignmentbeam} becomes
\begin{equation}
\mathcal{U}(x, z) = \mathrm{E}_{0} \mathcal{U}_{n}(x, z) e^{\mathrm{i} \omega t}
\end{equation}
The field after the electro-optic beam deflector modulator (EOBD) becomes~\cite{Fulda:17}
\begin{equation}
\begin{aligned}
\mathcal{U}^{\text{EOBD}} &\approx \mathrm{E}_{0}\mathcal{U}_{n} e^{\mathrm{i} \omega t}
+ \mathrm{i} \frac{m_{\alpha}}{2\Theta_{m}}\mathrm{E}_{0} \bigg(\sqrt{n+1} \mathcal{U}_{n+1} e^{- \mathrm{i}\Psi_{1}} \\
&+ \sqrt{n} \mathcal{U}_{n-1} e^{\mathrm{i}\Psi_{1}}\bigg) \Big(e^{\mathrm{i}\left(\omega+\Omega\right) t} + e^{\mathrm{i}\left(\omega-\Omega\right)t}\Big)
\label{equ:input_jitter}
\end{aligned}
\end{equation}
where $m_{\alpha}$ is the RF jitter modulation depth, $\Theta_{m}$ is the far-field beam divergence angle at the tilt-modulator location, and $\Psi_{1}$ is the accumulated Gouy phase from the modulator to the cavity waist. 

\subsubsection{RFJ: tilt}
Let us look at the tilt degree of freedom. Upon applying a tilt of angle $\alpha$ between the beam coming out of the EOBD and the cavity optical axis at the cavity waist, Eq.~\ref{equ:input_jitter} becomes
\begin{widetext}
\begin{equation}
\begin{aligned}
\mathcal{U}^{\alpha} &\approx 
\mathrm{E}_{0}\mathcal{U}_{n} e^{\mathrm{i} \omega t}
+ \mathrm{i} \frac{\alpha}{\Theta_{c}}\mathrm{E}_{0}\bigg(\sqrt{n+1} \mathcal{U}_{n+1} e^{- \mathrm{i}\Psi_{2}} + \sqrt{n} \mathcal{U}_{n-1} e^{\mathrm{i}\Psi_{2}}\bigg)
+ \mathrm{i} \frac{m_{\alpha}}{2\Theta_{m}} \bigg[
\sqrt{n+1} \Bigg(\mathcal{U}_{n+1} + \mathrm{i} \frac{\alpha}{\Theta_{c}} \Big(\sqrt{n+2} \mathcal{U}_{n+2}e^{- \mathrm{i}\Psi_{2}} + \sqrt{n+1} \mathcal{U}_{n} e^{\mathrm{i}\Psi_{2}}\Big) \Bigg)\\
&\times e^{- \mathrm{i}\Psi_{1}} + \sqrt{n} \Bigg(\mathcal{U}_{n-1} 
+ \mathrm{i} \frac{\alpha}{\Theta_{c}} \Big(\sqrt{n} \mathcal{U}_{n} e^{- \mathrm{i}\Psi_{2}} + \sqrt{n-1} \mathcal{U}_{n-2} e^{\mathrm{i}\Psi_{2}}\Big)\Bigg) e^{\mathrm{i}\Psi_{1}}\bigg] 
\left(e^{\mathrm{i}(\omega+\Omega) t} + e^{\mathrm{i}(\omega-\Omega) t}\right)
\end{aligned}
\label{equ:input_tilt}
\end{equation}
\end{widetext}
where $\Psi_{2}$ is the accumulated Gouy phase from the cavity waist to the PD. We can safely ignore $\Psi_{2}$ from now on since single-element photodetectors can only detect the beat between the same mode sidebands (such as $\mathrm{HG}_{n+1,0}$ and $\mathrm{HG}_{n+1,0}$, $\mathrm{HG}_{n-1,0}$ and $\mathrm{HG}_{n-1,0}$) assuming the aperture of the PD is much larger than the beam size. These same mode sidebands have the same accumulated Gouy phase $\Psi_{2}$. This overall phase factor however has no contribution to the photocurrent $\mathrm{E}\cdot\mathrm{E}^{*}$ as it cancels. Later on we will also see the extra accumulated Gouy phase $\Psi_{1}$ for the sideband modes can also be `absorbed' into an overall demodulation phase and does not concern us. 

In order to extract the error signals in reflection of the cavity, we need to multiply each term in Eq.~\ref{equ:input_tilt} with a suitable reflectance function $F(\omega, n)$ of the cavity to obtain $\text{E}_{\text{refl}}$. In general the cavity reflectance function $F(\omega, n)$ in Eq.~\ref{equ:fwn} is complex. However, in the high-finesse, completely over-coupled cavity case, all resonant components have the reflectance function of -1 whereas non-resonant field components have the reflectance function of 1. RFJ scheme relies on modulating at the higher-order mode difference frequency, which makes the upper mode in the upper sideband and the lower mode in the lower sideband resonant in the optical cavity. As a result, we have $F(\omega, n) = F(\omega+\Omega,n+1) = F(\omega-\Omega,n-1) = -1$ while the rest reflectance values for the non-resonant field components, such as $F(\omega, n\pm1)$, are all 1. The photocurrent signal on the single-element photodetector, which is defined as
\begin{equation}
\mathrm{I}_{PD}= \int_{-\infty}^{\infty} dx \, E_{\mathrm{refl}} \cdot E_{\mathrm{refl}}^{*}
\label{equ:singlePD}
\end{equation}
becomes
\begin{equation}
\begin{aligned}
\mathrm{I}_{PD} = \frac{ m_{\alpha} \alpha}{\Theta_{m} \Theta_{c}} \mathrm{E}_{0}^{2} \left(2n+1\right) \left(e^{\mathrm{i} \Psi_{1}} e^{\mathrm{i} \Omega t}  +  e^{-\mathrm{i} \Psi_{1}} e^{-\mathrm{i} \Omega t}\right)
\end{aligned}
\label{equ:photocurrentRFJ}
\end{equation}
where we again have assumed a responsivity of 1 A/W, and used the orthonormal condition on a single-element photodiode in Eq.~\ref{equ:beatcoe}
\begin{equation}
\int_{-\infty}^{\infty} dx \, \mathcal{U}_{n} \cdot \mathcal{U}_{n^{\prime}}  = \delta_{n,n^{\prime}} 
\label{equ:orthbeat}
\end{equation}

Now let us perform the mathematical operations equivalent to the demodulation process. With demodulation, we multiply the above photocurrent signal (after being converted to a voltage
signal by a trans-impedance stage with a gain of $1 V/W$) by a local oscillator $\cos(\Omega t + \delta \phi) = \frac{1}{2} \left(e^{\mathrm{i}(\Omega t + \delta \phi)} + e^{-\mathrm{i}(\Omega t + \delta \phi)}\right)$ and then extract the DC terms with a low-pass filtering.
\begin{equation}
\begin{aligned}
&\mathrm{V}^{\Omega}_{\mathrm{RFJ}} = \mathrm{I}_{PD} \times \cos(\Omega t + \delta \phi) \\
&=  \frac{ m_{\alpha} \alpha \mathrm{E}_{0}^{2} (2n+1)\Big(e^{\mathrm{i} \Psi_{1}} e^{\mathrm{i} \Omega t}  +  e^{-\mathrm{i} \Psi_{1}} e^{-\mathrm{i} \Omega t}\Big)}{\Theta_{m} \Theta_{c}} \times \frac{e^{\mathrm{i}(\Omega t + \delta \phi)} + e^{-\mathrm{i}(\Omega t + \delta \phi)}}{2} \\
&= \frac{ m_{\alpha} \alpha \mathrm{E}_{0}^{2} (2n+1)}{2\Theta_{m} \Theta_{c}}\left(e^{\mathrm{i} (\Psi_{1} - \delta \phi)} + e^{\mathrm{i} (-\Psi_{1} + \delta \phi)}\right) + \left(\text{terms in } 2\Omega\right)
\end{aligned}
\label{equ:photocurrentRFJdemod}
\end{equation}
The DC terms can be maximized by setting the overall demodulation phase offset $\delta \phi$ to be $\Psi_{1}$. We have thus seen that the difference in the accumulated Gouy phases in the upper and lower sidebands does not concern us here if we adjust the overall demodulation phase correspondingly (and we know how much exactly we should adjust). Demodulating with demodulation phase set to be $\Psi_{1}$ (i.e. I-phase
demodulation), we obtain the error signal
\begin{equation}
\begin{aligned}
\mathrm{V}^{\Omega,\, \mathrm{I}}_{\mathrm{RFJ}} = \frac{ m_{\alpha} \alpha}{\Theta_{m} \Theta_{c}} \mathrm{E}_{0}^{2} (2n+1)
\end{aligned}
\label{equ:jitter_tilt}
\end{equation}
The relative sensing gain then is simply
\begin{equation}
\begin{aligned}
 \Sigma_{n} = 2n+1
\end{aligned}
\label{equ:rfjsensinggain}
\end{equation}

\subsubsection{RFJ: lateral offset}

Now let us consider the lateral offset error signal. Upon applying a lateral displacement of amount $\delta x_{0}$, Eq.~\ref{equ:input_jitter} becomes~\cite{Tao:21}
\begin{widetext}
\begin{equation}
\begin{aligned}
\mathcal{U}^{\delta x_{0}}&\approx 
\mathrm{E}_{0}\mathcal{U}_{n} e^{\mathrm{i} \omega t}
- \frac{\delta x_{0}}{w_{0}}\mathrm{E}_{0}e^{\mathrm{i} \omega t} \bigg(\sqrt{n+1} \mathcal{U}_{n+1} e^{- \mathrm{i}\Psi_{2}} - \sqrt{n} \mathcal{U}_{n-1} e^{\mathrm{i}\Psi_{2}}\bigg) + \mathrm{i} \frac{m_{\alpha}}{2\Theta_{m}}\mathrm{E}_{0} \bigg[
\sqrt{n+1} \Bigg(\mathcal{U}_{n+1} -  \frac{\delta x_{0}}{w_{0}} \Big(\sqrt{n+2} \mathcal{U}_{n+2} e^{- \mathrm{i}\Psi_{2}} \\
& - \sqrt{n+1} \mathcal{U}_{n} e^{\mathrm{i}\Psi_{2}}\Big) \Bigg) e^{- \mathrm{i}\Psi_{1}} 
+ \sqrt{n} \Bigg(\mathcal{U}_{n-1} - \frac{\delta x_{0}}{w_{0}} \Big(\sqrt{n} \mathcal{U}_{n} e^{- \mathrm{i}\Psi_{2}}- \sqrt{n-1} \mathcal{U}_{n-2} e^{\mathrm{i}\Psi_{2}}\Big)\Bigg) e^{\mathrm{i}\Psi_{1}}\bigg] 
\left(e^{\mathrm{i}(\omega+\Omega) t} + e^{\mathrm{i}(\omega-\Omega) t}\right)
\end{aligned}
\end{equation}
\end{widetext}
We can again ignore $\Psi_{2}$ since the field components that have contributions to the single-element photodetector signal have the same accumulated Gouy phase $\Psi_{2}$. This overall phase factor drops out and has no contribution to the photocurrent. Using the fact that $F(\omega, n) = F(\omega+\Omega,n+1) = F(\omega-\Omega,n-1) = -1$, and $F=1$ for all other terms for our high-finesse, completely over-coupled cavity, the single-element PD signal, as defined in Eq.~\ref{equ:singlePD}, becomes
\begin{equation}
\begin{aligned}
\mathrm{I}_{PD} = \frac{\mathrm{i} m_{\alpha}}{\Theta_{m}} \frac{\delta x_{0}}{w_{0}}\mathrm{E}_{0}^{2} (2n+1) \left(e^{\mathrm{i}\Psi_{1}}e^{\mathrm{i}\Omega t} - e^{-\mathrm{i}\Psi_{1}}e^{-\mathrm{i}\Omega t}\right)
\label{equ:error_offset}
\end{aligned}
\end{equation}
where we have used the orthonormal condition on a single-element photodiode in Eq.~\ref{equ:orthbeat}. For the demodulation process, similar to Eq.~\ref{equ:photocurrentRFJdemod}, we can adjust the overall demodulation phase to be $\Psi_{1}+\pi/2$ (a condition often known as quadrature-phase or Q-phase
demodulation) to extract the optimal error signal
\begin{equation}
\begin{aligned}
\mathrm{V}^{\Omega,\, \mathrm{Q}}_{\mathrm{RFJ}} = -\frac{ m_{\alpha} \delta x_{0}}{\Theta_{m} w_{0}} \mathrm{E}_{0}^{2} (2n+1)
\end{aligned}
\label{equ:jitter_offset}
\end{equation}
We thus see we can extract the two orthogonal alignment sensing error signals in Eqs.~\ref{equ:jitter_tilt} and~\ref{equ:jitter_offset} simultaneously with a single element photodetector with orthogonal
demodulation phases $\Psi_{1}$ and $\Psi_{1} + \pi/2$ respectively. The resulting error signals are proportional to each other, and they scale as $2n+1$ with $n$ being the mode order of the carrier $\text{HG}_{n,0}$, see Fig.~\ref{fig:approx1}. This linear dependence in $n$ comes from the fact that the ``effective'' modulation depth for the EOBD in Eq.~\ref{equ:input_jitter} scales roughly as $\sqrt{n}$. Beating it with the scattered modes caused by the static misalignment (whose amplitude also scales roughly as $\sqrt{n}$ as shown in Eq.~\ref{equ:input_tilt}) results in the linear dependence for the RFJ sensing gain here. We also conduct the corresponding \textsc{Finesse} simulation. As shown in blue in Fig.~\ref{fig:approx1}, they agree extremely well with the analytical results in orange.

We thus see that the alignment sensing error signals in RFJ scheme increase linearly with respect to the mode index, as shown in Fig.~\ref{fig:approx1}. This increase is even faster than the scaling relation in the traditional WFS technique in Eq.~\ref{equ:error_tilt_WFS}, due to the extra factor of $\sqrt{n}$ from the effective modulation depth of the EOBD for the jitter-based scheme.

\section{Mode-Mismatch Sensing}
\label{sec:modemismatch}
For the mode mismatch sensing, as illustrated in the bottom panels of Fig.~\ref{fig:alignmentModeMismatch}, we can’t reduce the problem to a single-axis case, so we will have to consider both $x$ and $y$ axes at the same time~\cite{Tao:21}. In this Section, we will calculate the mode mismatch sensing error signals with an arbitrary higher-order HG mode $\mathrm{HG}_{n,m}$ as the carrier in both the mode-converter sensing scheme (MCS) the more recently proposed radio-frequency lens modulation sensing scheme (RFL). For the sake of compactness and simplicity, we will consider the case where the mode mismatch occurs at the center of a symmetric cavity, which we make coincident with the origin of our coordinate system, i.e. $z_{0} = 0$. As a result, we can simplify the initial beam in Eq.~\ref{equ:initialbeam} as
\begin{equation}
E(x, y, z)=E_{0} \mathcal{U}_{n, m}(x,y,z) e^{\mathrm{i}\omega t}
\label{equ:modemismatchbeam}
\end{equation}
where $\mathcal{U}_{n,m}(x,y,z)$ is the transverse function representing the $\text{HG}_{nm}$ mode in Eq.~\ref{equ:HGnm}.

\subsection{Mode-Mismatch Sensing: Mode-converter Sensing (MCS)}
For the MCS scheme, we make use of a $\pi/2$ mode converter and two $45^\circ$-rotated quadrant photodetectors away from the cavity waist by $\pi/4$ and $\pi/2$ Gouy phases for the two orthogonal waist size and waist position mismatch degrees of freedom, respectively. This scheme was first introduced by Magaña-Sandoval \textit{et al.}~\cite{Maga_a_Sandoval_2019} in the case of $\mathrm{HG}_{0,0}$ mode mismatch sensing. In this section, however, we are going to extend this scheme to a generic $\mathrm{HG}_{n,m}$ mode mismatch sensing. Similar to the alignment error signal calculation, we start with applying a phase modulation with modulation index m at a frequency $\Omega$ to the carrier $\text{HG}_{n,m}$ mode. Keeping only the first-order sidebands, the field becomes
\begin{equation}
E=E_{0} \mathcal{U}_{n,m} e^{\mathrm{i} \omega t}\left(1+\mathrm{i} \frac{m}{2}\left(e^{-\mathrm{i} \Omega t}+e^{\mathrm{i} \Omega t}\right)\right)
\end{equation}

\subsubsection{MCS: waist size mismatch}
Let us consider the waist size mismatch degree of freedom first. Upon application of a waist size mismatch between the input beam and the cavity eigenmode, the input beam becomes~\cite{Tao:21}
\begin{equation}
\begin{aligned}
\mathcal{U}_{n,m}^{\epsilon}(x, y, z) &\approx \mathcal{U}_{n,m} -\mathrm{i}\frac{\epsilon}{2} \Big( A_{n}\mathcal{U}_{n+2,m} + B_{n} \mathcal{U}_{n-2,m} \\
&+ A_{m}\mathcal{U}_{n,m+2} + B_{m}\mathcal{U}_{n,m-2}\Big)
\end{aligned}
\end{equation}
where $\epsilon = \frac{w}{w_{0}} - 1$ is the relative waist size mismatch, and $A_{n} = \sqrt{(n+2)(n+1)}$ and $B_{n} =\sqrt{n(n-1)}$. We have also used the fact that the accumulated Gouy phase from the cavity center to the QPD for each mode order is $\pi/4$ in the above equation. This introduces a factor of $e^{i 
\frac{\pi}{4}\cdot (\pm 2)} = \pm i$ for the upper and lower adjacent modes scattered from the original mode by two mode order as the result of mode mismatch. As illustrated above, in the mode mismatch sensing with WFS scheme we use a $\pi/2$ mode-converter, which causes an additional $\pi/2$ Gouy phase accumulation and thus a factor of $e^{i\pi/2}=i$ for each mode order in the focusing axis ($y$ axis) while the nonfocusing axis ($x$ axis) experiences normal Gouy phase accumulation~\cite{Maga_a_Sandoval_2019}. The above mode mismatched beam after passing through the mode converter becomes
\begin{equation}
\begin{aligned}
\mathcal{U}_{n,m}^{\epsilon}(x, y, z) \approx& \mathrm{i}^{m} \bigg(\mathcal{U}_{n,m} -\mathrm{i}\frac{\epsilon}{2} \Big( A_{n}\mathcal{U}_{n+2,m}+ B_{n} \mathcal{U}_{n-2,m} \\
& - A_{m}\mathcal{U}_{n,m+2} - B_{m}\mathcal{U}_{n,m-2}\Big)\bigg)
\end{aligned}
\label{equ:WS}
\end{equation}
where every mode order in the y direction accumulates one extra factor of $e^{i \pi/2} = i$ due to the $\pi/2$ mode-converter.

Similar to the alignment sensing case, we multiply each term in the incoming beam with the corresponding cavity reflectance function to get the reflected beam $\mathrm{E}_{\text{refl}}$. After making the assumption about a high-finesse and completely over-coupled cavity so that only $F(\omega, \mathrm{n+m}) = -1$ while $F=1$ for all other terms to simplify our analytical results, we obtain
\begin{widetext}
\begin{equation}
\begin{aligned}
\text E_{refl} &= \text E_{0} \mathrm{i}^{m} e^{\mathrm{i} \omega t} \Bigg(- \mathcal{U}_{n,m} + \frac{\mathrm{i} m}{2}\bigg(\mathcal{U}_{n,m} e^{\mathrm{i} \Omega t} + \mathcal{U}_{n,m} e^{-\mathrm{i} \Omega t}\bigg) - \frac{\mathrm{i} \epsilon}{2} \bigg( \mathcal{U}_{n+2,m} A_{n} - \mathcal{U}_{n,m+2} A_{m} + \mathcal{U}_{n-2,m} B_{n}- \mathcal{U}_{n,m-2} B_{m} \bigg) + \frac{\epsilon m}{4} \bigg( \mathcal{U}_{n+2,m}  A_{n} e^{\mathrm{i}\Omega t} \\
&+ \mathcal{U}_{n+2,m} A_{n} e^{-\mathrm{i}\Omega t} - \mathcal{U}_{n,m+2} A_{m} e^{\mathrm{i}\Omega t} - \mathcal{U}_{n,m+2} A_{m} e^{-\mathrm{i}\Omega t} + \mathcal{U}_{n-2,m} B_{n} e^{\mathrm{i}\Omega t} +\mathcal{U}_{n-2,m} B_{n} e^{-\mathrm{i}\Omega t} -  \mathcal{U}_{n,m-2} B_{m}e^{\mathrm{i}\Omega t}  - \mathcal{U}_{n,m-2} B_{m} e^{-\mathrm{i}\Omega t} \bigg) \Bigg)
\end{aligned}
\label{equ:refl}
\end{equation}
\end{widetext}
$E_{\mathrm{refl}}$ and $E_{\mathrm{refl}}^{*}$ have 15 terms each, so there are 225 terms to evaluate in the product $E_{\mathrm{refl}}E_{\mathrm{refl}}^{*}$ for the photocurrent. However, only terms that are odd with respect to $y = \pm x$ have contributions to the $45^\circ$-rotated QPD that we are using. We also know only terms that oscillate with the modulation frequency $\Omega$ have contribution to the photodetector signal after demodulation at the RF modulation frequency $\Omega$ and low-pass filtering. The photocurrent is then
\begin{equation}
\begin{aligned}
&\mathrm{I}_{QPD}=\mathrm{E}_{0}^{2}\epsilon m \left(e^{\mathrm{i} \Omega t} + e^{-\mathrm{i} \Omega t}\right)\Big(-A_{n}\beta_{(n,m),(n+2,m)} \\
&+ A_{m} \beta_{(n,m),(n,m+2)} -B_{n}\beta_{(n,m),(n-2,m)} + B_{m}\beta_{(n,m),(n,m-2)}\Big)
\label{equ:photocurrent_mismatch}
\end{aligned}
\end{equation}
where $\beta_{(n,m),(n^{\prime},m^{\prime})}$ are the beat coefficient at a $45^\circ$-rotated quadrant photodetector
\begin{equation}
\begin{aligned}
 &\beta_{(n,m),(n^{\prime},m^{\prime})} =  \bigg(\int_{0}^{\inf}dy \int_{-y}^{y}dx  + \int_{-\inf}^{0}dy \int_{y}^{-y}dx \\
 &- \int_{0}^{\inf}dx \int_{-x}^{x}dy - \int_{-\inf}^{0}dx \int_{x}^{-x}dy \bigg) \, \mathcal{U}_{n,m} \cdot \mathcal{U}_{n^{\prime},m^{\prime}}
\end{aligned}
\label{equ:beat_QPD}
\end{equation}
One can easily confirm that the beat coefficient $\beta_{(n,m),(n^{\prime},m^{\prime})}$ satisfies 
\begin{equation}
\begin{aligned}
\beta_{(n,m),(n^{\prime},m^{\prime})} = \beta_{(n^{\prime},m^{\prime}),(n,m)} = -\beta_{(m,n),(m^{\prime},n^{\prime})}
\end{aligned}
\label{equ:beatrelation}
\end{equation}

Demodulating at phase $\phi=0$ we obtain the I-phase signal
\begin{equation}
\begin{aligned}
\mathrm{V}^{\Omega,\, \mathrm{QPD}_{1}}_{\mathrm{MCS}} &= \mathrm{E}_{0}^{2}\epsilon m\Big(-A_{n}\beta_{(n,m),(n+2,m)} + A_{m}\beta_{(n,m),(n,m+2)} \\
&-B_{n}\beta_{(n,m),(n-2,m)} + B_{m}\beta_{(n,m),(n,m-2)}\Big)
\end{aligned}
\label{equ:error_w0}
\end{equation}

It is sometimes common to use the Rayleigh range $z_{R} = \frac{\pi w^{2}}{\lambda} \approx z_{R0} + \frac{2\pi w_{0}^{2}}{\lambda}\cdot \epsilon$ to characterize the amount of waist size mismatch. With the Rayleigh range mismatch $\delta z_{R} = \frac{2\pi w_{0}^{2}}{\lambda}\cdot \epsilon$ as the expansion parameter, we can rewrite the error signal as
\begin{equation}
\begin{aligned}
\mathrm{V}^{\Omega,\, \mathrm{QPD}_{1}}_{\mathrm{MCS}}  &= \mathrm{E}_{0}^{2} \frac{\lambda m}{2\pi \omega_{0}^{2}} \delta z_{R} \bigg(- A_{n} \beta_{(n,m),(n+2,m)} + A_{m} \beta_{(n,m),(n,m+2)} \\
&- B_{n} \beta_{(n,m),(n-2,m)} + B_{m} \beta_{(n,m),(n,m-2)} \bigg)
\end{aligned}
\label{equ:error_zR}
\end{equation}

We define the relative sensing gain $\Omega_{n,m}$ as the ratio of the error signal for the higher-order mode $\mathrm{HG}_{n, m}$ and the fundamental mode $\mathrm{HG}_{0,0}$
\begin{equation}
\begin{aligned}
\Omega_{n,m} &\equiv \Big(- A_{n} \beta_{(n,m),(n+2,m)} + A_{m} \beta_{(n,m),(n,m+2)} - B_{n} \beta_{(n,m),(n-2,m)} \\
&+ B_{m} \beta_{(n,m),(n,m-2)}\Big)/\left(2\sqrt{2} \beta_{(0,0), (0,2)}\right)
\label{equ:gainmcs}
\end{aligned}
\end{equation}

To make it more precise, let us consider the case where $n=m$, i.e. we are considering symmetric higher-order HG modes such as $\mathrm{HG}_{0,0}$, $\mathrm{HG}_{1,1}$, $\mathrm{HG}_{2,2}$, $\mathrm{HG}_{3,3}$, etc. As a result, using the properties of $\beta_{(n,m),(n^{\prime}, m^{\prime})}$ in Eq.~\ref{equ:beatrelation}, we can simplify the relative sensing gain to 
\begin{equation}
\begin{aligned}
\Omega_{n,n} =  \bigg(A_{n} \beta_{(n,n),(n,n+2)}  + B_{n} \beta_{(n,n),(n,n-2)}\bigg)/\left(\sqrt{2} \beta_{(0,0), (0,2)}\right)
\label{equ:approxgain1}
\end{aligned}
\end{equation}
We thus have an increasing mode mismatch sensing gain for higher-order modes.

\subsubsection{MCS: waist position mismatch}

Now let's look at the waist position mismatch error signal. The beam after the waist position displacement $\delta z_{0}$, according to the results in paper~\cite{Tao:21}, is
\begin{equation}
\begin{aligned}
\mathcal{U}_{n,m}^{\delta z_{0}}&(x, y, z) \approx \mathcal{U}_{n,m} - \mathrm{i} \frac{\lambda \delta z_{0}}{4 \pi w_{0}^{2}} \Big( A_{n}\mathcal{U}_{n+2,m} + B_{n} \mathcal{U}_{n-2,m} \\
&+ A_{m} \mathcal{U}_{n,m+2} + B_{m} \mathcal{U}_{n,m-2} - (C_{n}+C_{m})\cdot \mathcal{U}_{n,m}\Big)
\end{aligned}
\end{equation}
where $C_{n}=2n+1$, and we have used the fact that the accumulated Gouy phase from the cavity waist to the QPD in this path for each mode order is $\pi/2$. And this brings an extra factor of $e^{i \frac{\pi}{2} \cdot (\pm2)} = -1$ for the offset modes. After the $\pi/2$ mode-converter, the beam becomes
\begin{equation}
\begin{aligned}
\mathcal{U}_{n,m}^{\delta z_{0}}&(x, y, z) = \mathrm{i}^{m} \bigg(\mathcal{U}_{n,m} - \mathrm{i} \frac{\lambda \delta z_{0}}{4 \pi w_{0}^{2}} \Big( A_{n} \mathcal{U}_{n+2,m} + B_{n}\mathcal{U}_{n-2,m} \\
&- A_{m}\mathcal{U}_{n,m+2} - B_{m} \mathcal{U}_{n,m-2} - (C_{n}+C_{m})\cdot \mathcal{U}_{n,m}\Big)\bigg)
\end{aligned}
\label{equ:WP}
\end{equation}

Comparing Eqs.~\ref{equ:WS} and~\ref{equ:WP} we see that we only need to track two changes going from the waist size mismatch error signal result in Eq.~\ref{equ:error_w0} to the unknown waist position mismatch error signal we are calculating: (a) we need to replace the $\mathcal{U}_{n,m}$ coefficient from 1 to $(1+\mathrm{i}\frac{\epsilon}{2} (C_{n}+C_{m}))$; (b) we need to replace the mode mismatch parameter $\epsilon$ with $\frac{\lambda \delta z_{0}}{2 \pi w_{0}^{2}}$. The additional term resulting from Replacement (a) is proportional to $\mathcal{U}_{n,m}$ and thus has to beat with the upper or lower modes, such as $\mathcal{U}_{n\pm 2,m}$ and $\mathcal{U}_{n,m\pm2}$ to have a non-vanishing beat signal. This additional beat signal however is of second order in $\epsilon$ and thus does not affect the linear error signal, as can be seen from Eq.~\ref{equ:refl}. As a result, Replacement (a) has zero net contribution to the photocurrent and subsequently the error signal up to the linear order. Now to obtain the waist position mismatch error signal, one simply needs to make the Replacement (b), which results in
\begin{equation}
\begin{aligned}
\mathrm{V}^{\Omega,\, \mathrm{QPD}_{2}}_{\mathrm{MCS}}  &= \mathrm{E}_{0}^{2} \frac{\lambda \delta z_{0}}{2\pi w_{0}^{2}} m \Big(- A_{n} \beta_{(n,m),(n+2,m)} + A_{m} \beta_{(n,m),(n,m+2)} \\
&- B_{n} \beta_{(n,m),(n-2,m)} + B_{m} \beta_{(n,m),(n,m-2)} \Big)
\end{aligned}
\label{equ:error_z0}
\end{equation}
which interestingly has the same slopes as the waist size mismatch error signal in Eq.~\ref{equ:error_zR}. And we have the same sensing gain as in the waist size sensing case in Eq.~\ref{equ:gainmcs}.

Similar to the alignment WFS case, we want to approximate our exact results in Eqs.~\ref{equ:error_zR} and~\ref{equ:error_z0} to have a better sense of the scaling relation of the sensing gain with respect to the carrier's mode order. As in the alignment sensing case, we can simplify Eq.~\ref{equ:approxgain1} by replacing the complicated higher-order beat coefficients (excluding $\beta_{(0,0), (0,2)}$) with a constant value of 0.41. For more details see~\ref{app:beatcoeffs}.

\begin{equation}
\begin{aligned}
\Omega_{n,n} \approx  0.41\cdot \left(\sqrt{(n+1)(n+2)} + \sqrt{n(n-1)}\right)/\left(\sqrt{2} \beta_{(0,0), (0,2)}\right)
\label{equ:approxgain2}
\end{aligned}
\end{equation}
where we have used the definitions for $A_{n}$ and $B_{n}$. 

\begin{figure}[htbp]
    \centering
    \includegraphics[width=\linewidth]{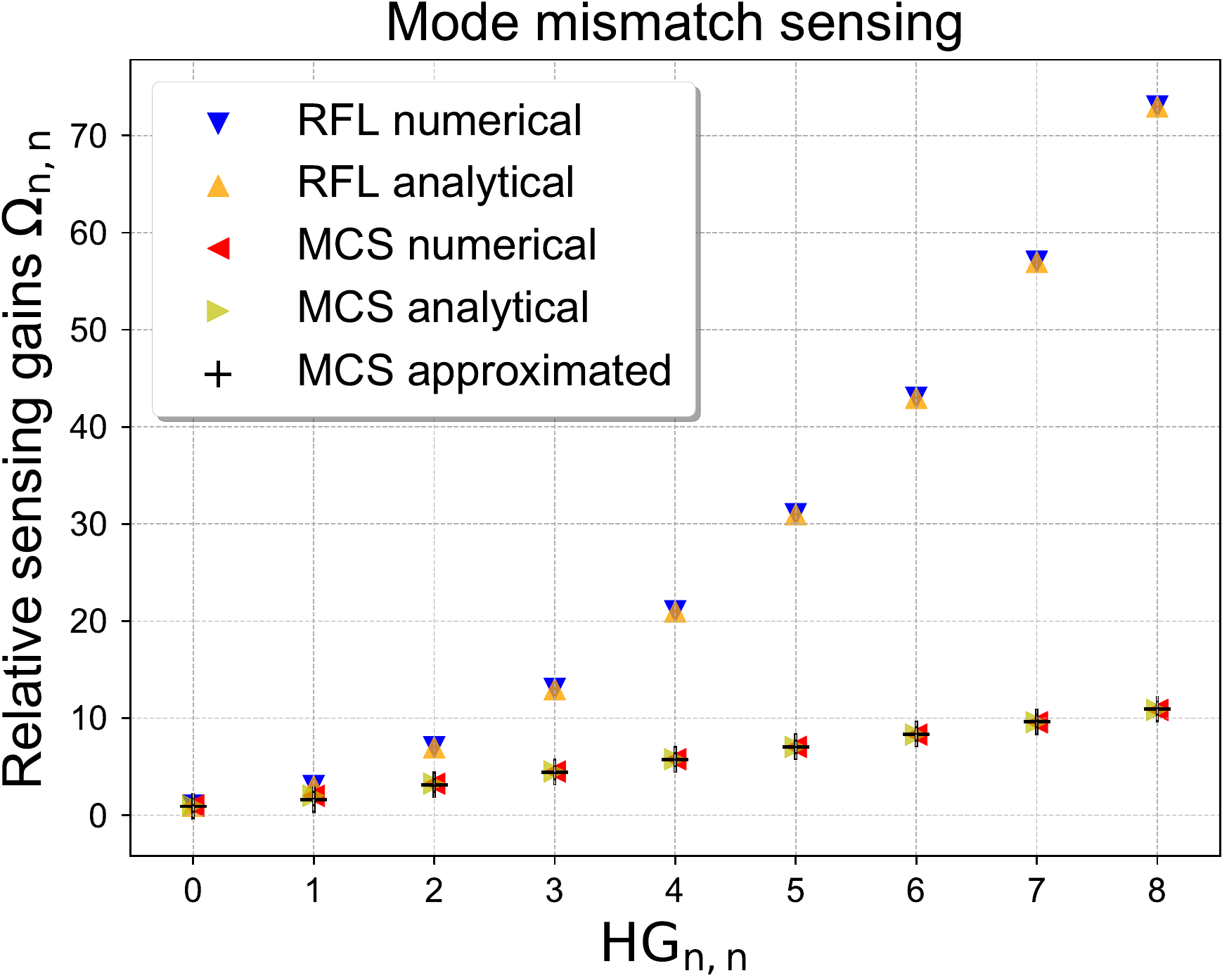}
    \caption{Relative mode mismatch sensing gain $\Omega_{n,n}$ in MCS and RFL sensing schemes for symmetric higher-order $\mathrm{HG}_{n, n}$ modes. The analytical sensing gains in Eqs.~\ref{equ:approxgain1} and~\ref{equ:rflsensinggain} and the approximated sensing gain in Eq.~\ref{equ:approxgain2} are shown. The corresponding \textsc{Finesse} simulation results are included as well for comparison.}
    \label{fig:approx2}
\end{figure}

This approximation is also quite close to the exact results, as shown in Fig.~\ref{fig:approx2}. We thus conclude that the mode mismatch sensing gain for higher order $\mathrm{HG}_{n,n}$ modes in this mode-converter and QPD sensing scheme scales approximately as $\sqrt{(n+1)(n+2)} + \sqrt{n(n-1)}$. We also conducted the corresponding \textsc{Finesse} simulation. As shown in black in Fig.~\ref{fig:approx2}, they agree extremely well with the exact analytical results in yellow.

\subsection{Mode-Mismatch Sensing: Radio-Frequency Lens Sensing (RFL)}
For the RFL mode mismatch sensing scheme, we use an electro-optic lens (EOL) device to impose RF lens modulation sidebands separated from the carrier frequency by \textit{twice} the higher-order mode difference frequency of the optical cavity, since we need the second order upper modes and lower modes generated from mode mismatch to resonate in the cavity. The scheme requires demodulating the beat signal between the RF lens modulation induced offset mode sidebands and the mode-mismatch-induced carrier frequency offset modes on a single element photodiode in the orthogonal demodulation phases. This can produce simultaneously linear error signals for the waist position and waist size mismatch between the input beam and the cavity eigenmode.

Similar to the mode-converter mode mismatch sensing scheme, we consider a generic $\text{HG}_{n,m}$ mode as the carrier in Eq.~\ref{equ:modemismatchbeam}
\begin{equation}
\mathcal{U}(x,y,z) = \mathrm{E}_{0}\mathcal{U}_{n,m}(x,y,z) e^{\mathrm{i} \omega t}
\end{equation}
We pass the carrier through an EOL device to modulate the curvature of the beam $S$. If we consider a sinusoidal oscillation in the wavefront curvature of the beam
\begin{equation}
S = m_{S} \cos(\Omega t)
\end{equation}
where $m_{S}$ is the RF lens modulation depth. The input beam after the EOL modulator becomes~\cite{Tao:21}
\begin{equation}
\begin{aligned}
\mathcal{U}^{\text{EOL}}&\approx \mathrm{E}_{0}\mathcal{U}_{n,m} e^{\mathrm{i} \omega t}
- \mathrm{i} \frac{k w_{0}^{2} m_{S}}{16} \mathrm{E}_{0} \bigg(A_{n} \mathcal{U}_{n+2,m} e^{- 2\mathrm{i}\Psi_{1}} \\
&+ B_{n} \mathcal{U}_{n-2,m} e^{2\mathrm{i}\Psi_{1}} + A_{m} \mathcal{U}_{n,m+2} e^{- 2\mathrm{i}\Psi_{1}} + B_{m} \mathcal{U}_{n,m-2} e^{2\mathrm{i}\Psi_{1}}
\\
&+ (C_{n}+C_{m})\cdot \mathcal{U}_{n,m}\bigg) \left(e^{\mathrm{i}(\omega+\Omega) t} + e^{\mathrm{i}(\omega-\Omega) t}\right)
\label{equ:input_lens}
\end{aligned}
\end{equation}
keeping only the first order terms, where $A_{n} = \sqrt{(n+2)(n+1)}$ and $B_{n} =\sqrt{n(n-1)}$.

\subsubsection{RFL: waist size mismatch}
Now let us consider the RFL waist size mismatch sensing error signal. Upon application of waist size mismatch $\epsilon = \frac{w}{w_{0}} - 1$, Eq.~\ref{equ:input_lens} becomes
\begin{widetext}
\begin{equation}
\begin{aligned}
\mathcal{U}^{\epsilon}&\approx 
\mathrm{E}_{0}\mathcal{U}_{n,m}e^{\mathrm{i} \omega t} + \mathrm{E}_{0}\frac{\epsilon}{2} \left(A_{n}\mathcal{U}_{n+2,m} - B_{n}\mathcal{U}_{n-2,m} + A_{m}\mathcal{U}_{n,m+2} - B_{m}\mathcal{U}_{n,m-2}\right)e^{\mathrm{i} \omega t} - \mathrm{i} \frac{k w_{0}^{2} m_{S}}{16}\mathrm{E}_{0} \bigg(A_{n} 
\Big(\mathcal{U}_{n+2,m} + \frac{\epsilon}{2} \big(A_{n+2} \mathcal{U}_{n+4,m} \\
&- B_{n+2}\mathcal{U}_{n,m} + A_{m}\mathcal{U}_{n+2,m+2} - B_{m}\mathcal{U}_{n+2,m-2}\big)\Big) e^{- 2\mathrm{i}\Psi_{1}} + B_{n} \Big(\mathcal{U}_{n-2,m} + \frac{\epsilon}{2} \big(A_{n-2}\mathcal{U}_{n,m} - B_{n-2}\mathcal{U}_{n-4,m} + A_{m}\mathcal{U}_{n-2,m+2} - B_{m}\mathcal{U}_{n-2,m-2}\big)\Big)
 e^{2\mathrm{i}\Psi_{1}} \\
 &+ 
A_{m} \Big(\mathcal{U}_{n,m+2} + \frac{\epsilon}{2} \big(A_{n}\mathcal{U}_{n+2,m+2} - B_{n}\mathcal{U}_{n-2,m+2} + A_{m+2}\mathcal{U}_{n,m+4} - B_{m+2}\mathcal{U}_{n,m}\big)\Big) e^{- 2\mathrm{i}\Psi_{1}} + B_{m} \Big(\mathcal{U}_{n,m-2} + \frac{\epsilon}{2} \big(A_{n}\mathcal{U}_{n+2,m-2} - B_{n}\mathcal{U}_{n-2,m-2}\\
&+ A_{m-2}\mathcal{U}_{n,m} - B_{m-2}\mathcal{U}_{n,m-4}\big)\Big) e^{2\mathrm{i}\Psi_{1}}
+ \left(C_{n}+C_{m}\right)
\Big(\mathcal{U}_{n,m} + \frac{\epsilon}{2} \big(A_{n}\mathcal{U}_{n+2,m} - B_{n}\mathcal{U}_{n-2,m} + A_{m}\mathcal{U}_{n,m+2} - B_{m}\mathcal{U}_{n,m-2}\big)\Big)
\bigg)\left(e^{\mathrm{i}(\omega+\Omega) t} + e^{\mathrm{i}(\omega-\Omega) t}\right)
\end{aligned}
\label{equ:rflfield}
\end{equation}
\end{widetext}

where similar to RFJ we have omitted the accumulated Gouy phase difference $\Psi_{2}$ from the cavity waist to the single element PD because it does not affect the PD signal. The photocurrent on the single-element photodetector is defined as
\begin{equation}
\mathrm{I}_{PD}= \int_{-\infty}^{\infty}\int_{-\infty}^{\infty} dx dy \, E_{\mathrm{refl}} \cdot E_{\mathrm{refl}}^{*}
\end{equation}
where the reflected field $E_{\mathrm{refl}}$ is obtained by multiplying each term in Eq.~\ref{equ:rflfield} with a suitable reflectance function $F(\omega, n)$. Using the fact that $F(\omega, n+m) = F(\omega+\Omega,n+m+2) = F(\omega-\Omega,n+m-2) = -1$, and $F=1$ for all other terms for completely over-coupled cavity, the photocurrent becomes
\begin{equation}
\begin{aligned}
\mathrm{I}_{PD} = \frac{\mathrm{i}k w_{0}^{2} m_{S} \epsilon \mathrm{E}_{0}^{2}}{8} \left(n^{2}+m^{2}+n+m+2\right) \left(e^{2\mathrm{i}\Psi_{1}} e^{\mathrm{i}\Omega t} - e^{-2\mathrm{i}\Psi_{1}} e^{-\mathrm{i}\Omega t}\right)
\label{equ:error_ws}
\end{aligned}
\end{equation}
where we have used the orthonormal condition in Eq.~\ref{equ:beatcoe}.

Similar to the RFJ sensing scheme in Eq.~\ref{equ:photocurrentRFJdemod}, we can adjust the overall demodulation phase to be $2 \Psi_{1}+\pi/2$ (not $\Psi_{1}+\pi/2$) to eliminate the effect of accumulated Gouy phase difference between different sideband modes and maximize the demodulated signal. This Q-phase error signal is
\begin{equation}
\begin{aligned}
\mathrm{V}^{\Omega,\, \mathrm{Q}}_{\mathrm{RFL}} = -\frac{k w_{0}^{2} m_{S} \epsilon}{8} \mathrm{E}_{0}^{2} \left(n^{2}+m^{2}+n+m+2\right)
\end{aligned}
\label{equ:rfl_ws}
\end{equation} 
Using the Rayleigh range mismatch $\delta z_{R} = \frac{2\pi w_{0}^{2}}{\lambda}\cdot \epsilon$ as the expansion parameter, we can rewrite the error signal as
\begin{equation}
\begin{aligned}
\mathrm{V}^{\Omega,\, \mathrm{Q}}_{\mathrm{RFL}} = -\frac{m_{S} \mathrm{E}_{0}^{2}}{8} \delta z_{R} \left(n^{2}+m^{2}+n+m+2\right)
\end{aligned}
\label{equ:rfl_zr}
\end{equation}
Thus, the relative waist size sensing gain for $\mathrm{HG}_{n, n}$ mode in the RFL scheme is
\begin{equation}
\begin{aligned}
\Omega_{n,n} = n^2+n+1
\label{equ:rflsensinggain}
\end{aligned}
\end{equation}

\subsubsection{RFL: waist position mismatch}
Upon application of a waist position mismatch $\delta z_{0}$ between the input beam and the cavity eigenmode at the cavity waist, the input beam in Eq.~\ref{equ:input_lens} becomes
\begin{widetext}
\begin{equation}
\begin{aligned}
\mathcal{U}^{\delta z_{0}}&\approx 
\mathrm{E}_{0}\mathcal{U}_{n,m}e^{\mathrm{i} \omega t} + \mathrm{E}_{0}\frac{\mathrm{i} \lambda \delta z_{0}}{4\pi w_{0}^2}e^{\mathrm{i} \omega t} \Big(A_{n}\mathcal{U}_{n+2,m} + B_{n}\mathcal{U}_{n-2,m} + A_{m} \mathcal{U}_{n,m+2} + B_{m}\mathcal{U}_{n,m-2} + (C_{n}+C_{m})\mathcal{U}_{n,m}\Big)
- \mathrm{i} \frac{k w_{0}^{2} m_{S}}{16}\mathrm{E}_{0} \bigg(A_{n} \Big(\mathcal{U}_{n+2,m} \\
&+ \frac{\mathrm{i} \lambda \delta z_{0}}{4\pi w_{0}^2} \big(A_{n+2}\mathcal{U}_{n+4,m} + B_{n+2}\mathcal{U}_{n,m} + A_{m}\mathcal{U}_{n+2,m+2} + B_{m}\mathcal{U}_{n+2,m-2} + (C_{n+2}+C_{m})\mathcal{U}_{n+2,m}\big)\Big)
 e^{- 2\mathrm{i}\Psi_{1}} + B_{n} \Big(\mathcal{U}_{n-2,m} +\frac{\mathrm{i} \lambda \delta z_{0}}{4\pi w_{0}^2} \big(A_{n-2}\mathcal{U}_{n,m} \\
 &+ B_{n-2}\mathcal{U}_{n-4,m} + A_{m} \mathcal{U}_{n-2,m+2} + B_{m}\mathcal{U}_{n-2,m-2} + (C_{n-2}+C_{m})\mathcal{U}_{n-2,m}\big)\Big)
e^{2\mathrm{i}\Psi_{1}} +A_{m} \Big(\mathcal{U}_{n,m+2} +\frac{\mathrm{i} \lambda \delta z_{0}}{4\pi w_{0}^2} \big(A_{n}\mathcal{U}_{n+2,m+2} + B_{n}\mathcal{U}_{n-2,m+2} +\\
&A_{m+2} \mathcal{U}_{n,m+4} + B_{m+2}\mathcal{U}_{n,m} + (C_{n}+C_{m+2})\mathcal{U}_{n,m+2}\big)\Big)
e^{- 2\mathrm{i}\Psi_{1}} + B_{m} \Big(\mathcal{U}_{n,m-2} +\frac{\mathrm{i} \lambda \delta z_{0}}{4\pi w_{0}^2} \big(A_{n}\mathcal{U}_{n+2,m-2} + B_{n}\mathcal{U}_{n-2,m-2} + A_{m-2} \mathcal{U}_{n,m} \\
&+ B_{m-2}\mathcal{U}_{n,m-4} + (C_{n}+C_{m-2})\mathcal{U}_{n,m-2}\big)\Big) 
e^{2\mathrm{i}\Psi_{1}} + (C_{n}+C_{m}) \Big(\mathcal{U}_{n,m} +\frac{\mathrm{i} \lambda \delta z_{0}}{4\pi w_{0}^2} \big(A_{n}\mathcal{U}_{n+2,m} + B_{n}\mathcal{U}_{n-2,m} + A_{m}\mathcal{U}_{n,m+2} + B_{m}\mathcal{U}_{n,m-2} \\
&+ (C_{n}+C_{m})\mathcal{U}_{n,m}\big)\Big)
\bigg) \left(e^{\mathrm{i}(\omega+\Omega) t} + e^{\mathrm{i}(\omega-\Omega) t}\right)
\end{aligned}
\end{equation}
\end{widetext}
where similarly we have omitted the accumulated Gouy phase difference for different sidebands $\Psi_{2}$. We can similarly simplify the reflected field $E_{\mathrm{refl}}$ using the fact that $F(\omega, n+m) = F(\omega+\Omega,n+m+2) = F(\omega-\Omega,n+m-2) = -1$, and $F=1$ for all other terms for completely over-coupled cavity. The PD signal on a single-element PD becomes 
\begin{equation}
\begin{aligned}
\mathrm{I}_{PD} = -\frac{m_{S} \mathrm{E}_{0}^2}{8} \delta z_{0} \left(e^{2\mathrm{i}\Psi_{1}}e^{\mathrm{i}\Omega t} + e^{-2\mathrm{i}\Psi_{1}}e^{-\mathrm{i}\Omega t}\right) \left(n^{2}+m^{2}+n+m+2\right)
\label{equ:error_wp}
\end{aligned}
\end{equation}
Similarly, we can eliminate the effect of the accumulated Gouy phase difference between different sideband modes by adjusting the overall demodulation phase to be $2 \Psi_{1}$ (not $\Psi_{1}$). The result I-phase signal is
\begin{equation}
\begin{aligned}
\mathrm{V}^{\Omega,\, \mathrm{I}}_{\mathrm{RFL}} = -\frac{m_{S} \mathrm{E}_{0}^2}{8} \delta z_{0} \left(n^{2}+m^{2}+n+m+2\right)
\end{aligned}
\label{equ:rfl_wp}
\end{equation}
which interestingly has the same slope as the waist size mismatch error signal in Eq.~\ref{equ:rfl_zr}.

We have thus seen, similar to the alignment sensing with the RFJ scheme, we can extract the two orthogonal mode mismatch sensing error signals in Eqs.~\ref{equ:rfl_wp} and~\ref{equ:rfl_ws} with a single element photodetector in the orthogonal
demodulation phases. The resulting error signals in RFL scheme scale quadratically with respect to the mode order, as shown in Fig.~\ref{fig:approx2}. And this quadratic dependence comes from the fact that the ``effective'' modulation depth for the EOL in Eq.~\ref{equ:input_lens} scales roughly as $n$. Beating it with the scattered modes caused by the static mode mismatch (whose amplitude also scales roughly as $n$) results in the quadratic dependence for the RFL sensing gains, which scales even faster than the MCS technique.

\section{Conclusion}
\label{sec:conclusion}
We have offered a detailed analytical derivation for the first time for the alignment and mode mismatch sensing error signals with a generic higher-order HG mode as the carrier. Specifically, we have investigated the traditional wavefront sensing scheme (WFS) and the more recently proposed radio-frequency jitter modulation sensing scheme (RFJ) in Section~\ref{sec:alignment} for the alignment sensing; and for the mode mismatch sensing, we have studied the mode-converter sensing scheme (MCS) and the radio-frequency lens modulation sensing scheme (RFL) Section~\ref{sec:modemismatch}. We have seen analytically for instance the necessity for the $\pi/2$ accumulated Gouy phase difference between the two split photodetectors for the alignment WFS, and the necessity for the $\pi/2$ mode-converter and the $\pi/4$ accumulated Gouy phase difference between the two $45^{\circ}$ rotated QPDs for the mode mismatch sensing. We have also conducted the corresponding \textsc{Finesse} simulations for the alignment and mode mismatch sensing for higher-order HG modes, and the resultant relative sensing gains agree extremely well with the respective exact analytical results, as shown in Figs.~\ref{fig:approx1} and~\ref{fig:approx2}. 

\begin{table}[htbp]
\centering
\caption{Summary of the alignment and mode mismatch sensing gains, and induced power coupling losses in Ref.~\cite{Tao:21}.}
\begin{tabular}{c|c|c}
\hline \hline 
& \thead{Alignment} & \thead{Mode mismatch} \\
\hline
\makecell{\thead{Traditional \\ schemes}} & \makecell{WFS: \\ $\sqrt{n}+\sqrt{n+1}$} & \makecell{MCS: \\ $\sqrt{n(n-1)}+\sqrt{(n+1)(n+2)}$} \\
 \hline 
\makecell{\thead{Beam jitter \\ based schemes}} & RFJ: $2n+1$ & RFL: $n^2+n+1$ \\
 \hline 
\thead{Power loss} & $2n+1$ & $n^2+n+1$ \\
\hline \hline
\end{tabular}
\label{tab:summary}
\end{table}

We have shown that in all schemes the alignment and mode mismatch sensing signals are stronger for higher-order Hermite-Gauss modes, as summarised in Tab.~\ref{tab:summary} for the sensing gains. On the other hand, the shot noise at the photodetector in all of the analyzed sensing schemes is independent of the mode indices of the beam. Therefore the alignment or mode mismatch signal to shot noise ratio increases with mode indices exactly as derived for the sensing gains. In a shot noise-limited sensing regime, and a sensing noise-limited control loop, increasing mode index is therefore seen to reduce the residual alignment or mode mismatch error. This goes at least some way to nullify the downside of increasing mode indices in terms of the reduced alignment and mode mismatch tolerances as described in Refs.~\cite{Jones_2020, Tao:21}. In particular, the increases in the sensing gain with RFJ/L scheme \textit{exactly} match the decreases in the corresponding tolerances, as shown in the last two rows in Tab.~\ref{tab:summary}. This result shows that for these schemes the downside of using higher-order HG modes with respect to their suffering from excessive misalignment and mode mismatch-induced power losses could be eliminated. 

\section*{Acknowledgments}
This work was supported by National Science Foundation grants PHY-1806461 and PHY-2012021.

\appendix
\section*{Appendix: Beat coefficients}
\label{app:beatcoeffs}

\begin{figure}[htbp]
    \centering
    \includegraphics[width=\linewidth]{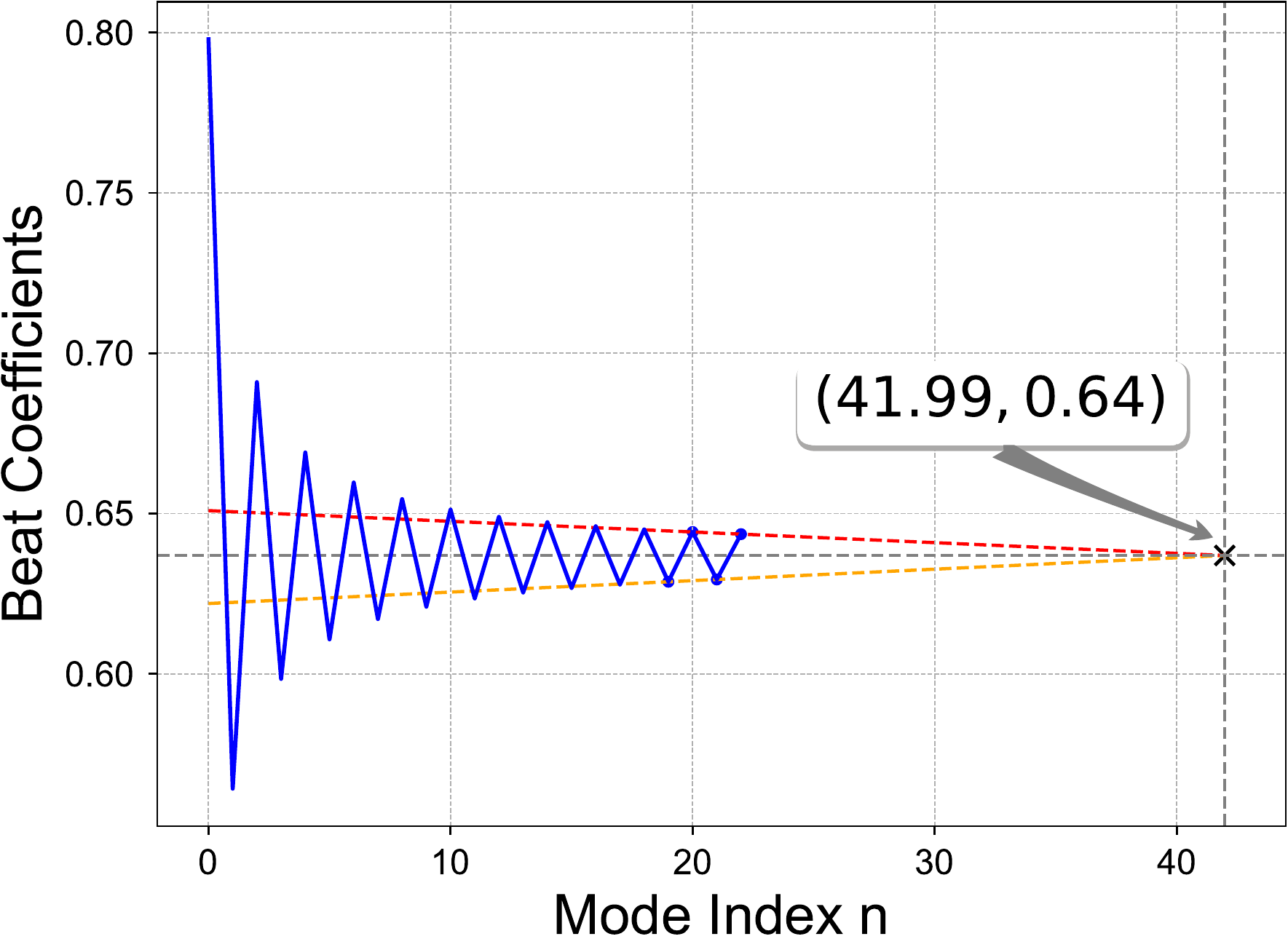}
    \caption{Beat coefficients $\beta_{n,n+1}$ on a split photodiode in blue. The two straight dashed lines connecting the four end points intersect at the one point with its y coordinate being around 0.64. }
    \label{fig:beatcoeff1}
\end{figure}

\begin{figure}[htbp]
    \centering
    \includegraphics[width=\linewidth]{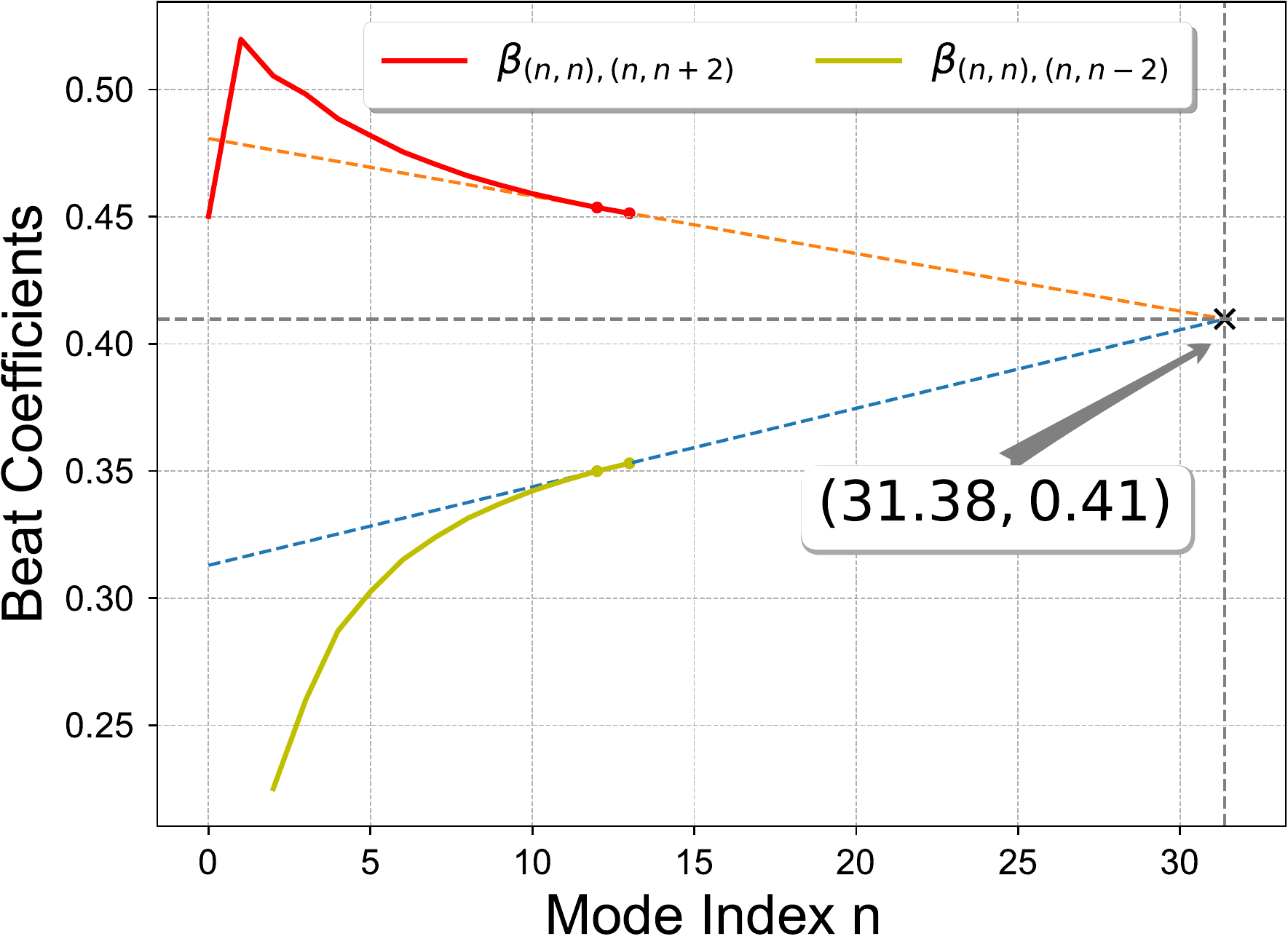}
    \caption{Beat coefficients $\beta_{(n,n),(n,n+2)}$ in red and $\beta_{(n,n),(n,n-2)}$ in yellow on a 45 degree rotated quadrant photodiode. They both converge to around 0.41 as n goes large. Notice that $\beta_{(n,n),(n,n-2)}$ have no definition when $n<2$, as shown in the yellow line.}
    \label{fig:beatcoeff2}
\end{figure}

Some example beat coefficients $\beta_{n,n+1}$ on a split photodetectors in Eq.~\ref{equ:beatsplit} and $\beta_{(n,n),(n,n+2)}$ and $\beta_{(n,n),(n,n-2)}$ on a quadrant photodetectors in Eq.~\ref{equ:beat_QPD} are shown in Figs.~\ref{fig:beatcoeff1} and~\ref{fig:beatcoeff2}. We can see that they both asymptotically approach to some fixed value around 0.64 and 0.41 respectively as n goes large, which are determined by linearly extrapolating from the end points and finding the intercepts. As a result, we can approximate the exact sensing gains in WFS and MCS in Eqs.~\ref{equ:gain1} and~\ref{equ:approxgain1} and simply the scaling relations by replacing the complicated higher-order beat coefficients with their rough asymptotic values, i.e. 0.64 for $\beta_{n,n+1}$ and 0.41 for $\beta_{(n,n),(n,n+2)}$ and $\beta_{(n,n),(n,n-2)}$.

\newpage
\bibliography{reference}

\end{document}